\documentstyle[aps,prd,preprint,epsfig]{revtex}

\begin{document}
\draft
\title{A universe in a global monopole}

\author{Katherine Benson\footnote{Electronic address: 
benson@physics.emory.edu} and 
Inyong Cho\footnote{Electronic address: 
cho@physics.emory.edu}}
\address{Department of Physics, Emory University,
Atlanta, Georgia 30322-2430, USA}

\date{\today}

\maketitle

\begin{abstract}

We investigate brane physics in a universe with an extra
dimensional global monopole and negative bulk cosmological
constant. The graviton zero mode is naturally divergent; we thus
invoke a physical cut-off to induce four dimensional gravity on a
brane at the monopole core.  Independently, the massive Kaluza-Klein
modes have naturally compactified extra dimensions, inducing a
discrete spectrum. This spectrum remains consistent with four
dimensional gravity on the brane, even for small mass gap.
Extra dimensional matter fields also induce four dimensional matter
fields on the brane, with the same Kaluza-Klein spectrum of excited
states.  We choose parameters to solve the hierarchy problem; that is,
to induce the observed hierarchy between particle and Planck scales in
the effective four dimensional universe.

\end{abstract}

\vspace{0.5in}

\section{Introduction}

In the 1920's Kaluza and Klein proposed that our universe may be
embedded in higher (five) dimensional spacetime~\cite{KK}. 
They sought to unify gravitational and electromagnetic forces in such
a proposal; specifically, to induce  four dimensional gauge interactions 
solely by the five dimensional geometry. 
Their fifth dimension was compactified and small in size, producing
negligible corrections to the traditional gravitational-force laws 
observed in four dimensions.

Recently the possibility of very large extra dimensions,
with interesting cosmological and
particle consequences, was proposed by Arkani-Hamed,
Dimopoulos and Dvali~\cite{ADD1}. 
In their proposal the compactified extra dimensions can be as
large as millimeter size without endangering four dimensional  
gravitational-force laws.
Through such large extra dimensions they solved the hierarchy
problem, by inducing a large effective 4D Planck mass from 
higher dimensional Planck mass,
through the relation
\begin{equation}
M_{Pl}^2 = V_nM_N^{2+n}\, .
\end{equation}
Here, $M_{Pl}$ and $M_N$ are the four and $N$ dimensional
Planck mass, $n$ is the number of extra dimensions, and
$V_n$ is the volume of the extra dimensions.

Soon after, Randall and Sundrum proposed that an extra
dimension can be infinitely large~\cite{RS}.  Their model
exploited a non-factorizable metric introduced by Rubakov and
Shaposhnikov~\cite{Rubakov2}, in which a ``warp factor''
decreases exponentially along the fifth coordinate.  Randall and
Sundrum realized this non-trivial geometry by embedding a 4D-matter
brane in 5D spacetime with a negative bulk cosmological constant.
They showed that the warp factor localizes massless gravitons on
the brane, reproducing the observed $1/r^2$ gravitational force, with
mild corrections due to a continuum of excited Kaluza-Klein (KK) graviton
states.  They also showed that, in integrating over the fifth
dimension to establish an effective four dimensional universe, the
warped metric can induce the observed hierarchy between the particle
and effective Planck scales.

Randall and Sundrum posited the brane, and intrinsically
four dimensional matter fields confined to it, from string theoretic
motivations. Subsequent work realized the four dimensional universe
more naturally, as a sub-manifold associated with topological defects formed by a
matter condensate in the extra dimensions. The defect solution
determines the warped metric, and binds both gravitons and matter
fields to a four dimensional internal space at the defect's core. Such
binding for matter fields is well-established (see, for example,
Refs.~\cite{Rubakov1,Rubakov2,Shifman} for 4D matter
bound to extra dimensional domain walls).  Solutions for warped
extra dimensional defect metrics appeared in Ref.~\cite{Vilenkin}; 
more complete explorations of solutions, showing bound $1/r^2$
four dimensional gravity plus corrections, and solved hierarchy
problems, appeared in
Refs.~\cite{Cohen,Shaposhnikov1,Shaposhnikov2}. Ghergetta, Roessl and
Shaposhnikov found gauged and global defects inducing theories with
massless gravitons and a continuum of Kaluza-Klein graviton states,
inducing small gravitational corrections. Cohen and Kaplan considered an
extra dimensional global string, and also found massless gravitons;
however, they found extra dimensions to be essentially compactified
by the existence of a singularity,
yielding a discrete spectrum of Kaluza-Klein modes 
with acceptable gravitational corrections.

In this paper we investigate the effective four dimensional
universe induced by an extra dimensional global monopole.  The
monopole forms in the extra three spatial dimensions 
of a 7D universe with negative 
cosmological constant, generating a warped
metric solution distinct from those considered previously.  Each
point in the extra transverse space corresponds to a 3D brane.  The
spacetime of this global monopole is singularity free in its
geometry~\cite{Vilenkin} unlike the global string model of Cohen
and Kaplan~\cite{Cohen}.  Therefore, the extra dimensional space
stretches without bound.

The volume of the extra dimensions in our model is infinite.  As
usual in models of infinite volume, the gravity zero mode is
not normalizable.  To normalize it, and achieve 4D gravity
on the brane, we introduce a cut-off. We regard this
cut-off as a natural element in our dynamical theory, measuring the
typical separation between global monopoles formed in the
dimension-reducing phase transition. We choose it so that the
effective Planck scale flows upward to generate the hierarchy.

Our model also contains a more inherent compactification of the
extra dimensions. Specifically, in transforming Einstein's equations
for Kaluza-Klein graviton modes into a Schrodinger-like equation, to
establish their mass spectrum, we find an effective radial variable
which is bounded. This necessarily yields a discrete graviton mass
spectrum. Unlike Cohen and Kaplan's bounding of the radial variable,
imposed to avoid a singularity, or our earlier introduction of a
cut-off, this compactification emerges necessarily from the form of
our nonsingular metric, in the presence of negative seven dimensional
cosmological constant. We find a model-dependent mass gap. Whether
this gap is small or large, we show the massive-graviton modes can
yield acceptable gravitational corrections.

We also investigate localization of 7D matter fields on
the brane. We find that 4D matter fields  are
induced on the brane, along with a tower of excited Kaluza-Klein
states whose mass gap and wave functions are identical to the graviton
modes.

The structure of this paper is as follows. In Sec.~II we present the
model.  In Sec.~III we discuss the graviton zero mode.  In Sec.~IV we
discuss massive Kaluza-Klein modes.  In Sec.~V matter fields are
discussed, and we conclude in Sec.~VI.

\section{The model}

We consider a global monopole formed in $n=3$ extra
transverse dimensions.
The general static form of the (4+3) dimensional metric with  
spherical symmetry in the extra three dimensions is
\begin{equation}
ds^2 = g_{MN}dx^Mdx^N
     = B(r)\bar{g}_{\mu\nu}dx^\mu dx^\nu +A(r)dr^2 +r^2d\Omega^2\,,
\label{eq=metric}
\end{equation}
where $\bar{g}_{\mu\nu}$ is the apparent 4D metric 
and we take  the mainly $+$ sign convention.\footnote{Note that the 
capital Roman index runs over all seven dimensions, while the
Greek index runs over our four longitudinal dimensions and the small Roman index
runs over the extra three transverse dimensions.}

The action is
\begin{equation}
{\cal S} = \int d^7x \sqrt{-g}\left(
{{\cal R}-2\Lambda \over 16\pi G_N} + {\cal L}_m \right)\,,
\label{eq=action}
\end{equation}
where $G_N$ is the 7D gravitational constant  
$G_N=1/8\pi M^{2+n}_N$, $\Lambda$ is the 7D cosmological 
constant,  and ${\cal L}_m$ is the Lagrangian of the monopole field
given by
\begin{equation}
{\cal L}_m = -{1\over 2}\partial_A\phi^a\partial^A\phi^a
	-V(\sqrt{\phi^a\phi^a})\,,
\end{equation}
where $\phi^a = \phi(r)\hat{x}^a$ is the scalar triplet,
$V={\lambda\over 4}(\phi^a\phi^a-\eta^2)^2$, 
and $\eta$ is the symmetry-breaking scale.

With the given action and metric Einstein's equation is
\begin{equation}
R_{MN} -{1\over 2}g_{MN}{\cal R}+\Lambda g_{MN}
= \kappa^2 T_{MN}\,,
\end{equation}
where $\kappa^2 = 8\pi G_N$ and 
$T_{MN}$ is the energy-momentum tensor given by
\begin{equation}
T_{MN} = \partial_M\phi^a \partial_N\phi^a + g_{MN}{\cal L}_m\,.
\end{equation}

Then each component of Einstein's equation becomes

\begin{eqnarray}
-G^\mu_\mu &=& {1\over A}\left[ -{3\over 2}{B'' \over B} 
+{3\over 4}{A'B' \over AB}
-3{B' \over Br} +{A' \over Ar}
+{A-1 \over r^2} +{1\over 4}{A \over B}\bar{R}^{(4)}\right] = 
\kappa^2\left[ {\phi'^2 \over 2A} +{\phi^2 \over r^2}+ V(\phi)
\right] +\Lambda \,, \nonumber\\
-G^r_r &=& {1\over A}\left[ -{3\over 2}\left({B'\over B}\right)^2 
-4{B' \over Br} 
+{A-1\over r^2} +{1\over 2}{A\over B}\bar{R}^{(4)}\right] = 
\kappa^2\left[ -{\phi'^2 \over 2A} +{\phi^2 \over r^2}+ V(\phi)
\right] +\Lambda \,, \label{einsteqns}\\
-G^{\theta_i}_{\theta_i} &=& 
{1\over A}\left[ -2{B'' \over B} 
-{1\over 2}\left({B'\over B}\right)^2
+{A'B' \over AB}
-2{B' \over Br} +{1\over 2}{A' \over Ar} 
+{1\over 2}{A\over B}\bar{R}^{(4)}\right] = 
\kappa^2\left[ {\phi'^2 \over 2A} + V(\phi)
\right] +\Lambda \,,\nonumber
\end{eqnarray}
where $\bar{R}^{(4)}$ is the 4D Ricci scalar induced by $\bar{g}_{\mu\nu}$. 
Note that we have exploited diagonality of the stress-energy tensor $T^M_N$ 
within the longitudinal subspace, to set  
$\bar{G}^\mu_\nu = - \bar{R}^{(4)}/4\ \delta^\mu_\nu$, 
where $\bar{G}^\mu_\nu$  is the 4D Einstein tensor induced by $\bar{g}_{\mu\nu}$.

The field equation for the scalar field is
\begin{equation}
{\phi''\over A} +{1\over A}\left( -{1\over 2}{A'\over A}
+2{B'\over B}+{2\over r}\right)\phi'
-{2 \over r^2}\phi -\lambda\phi (\phi^2-\eta^2) =0\,.
\label{eq=phi}
\end{equation}
The scalar field takes a hedgehog configuration 
with boundary conditions
\begin{equation}
\phi (0)=0\,,\quad \phi(\infty)=\eta\,.
\end{equation}

We first determine asymptotic behavior of solutions to these field
equations. Near the origin, the scalar field has linear dependence on $r$, 
\begin{equation}
\phi (r) \approx \phi_0 r\,.
\end{equation}
Imposing the regularity condition on the gravitational fields 
gives their asymptotic form at $r\approx 0$ as
\begin{eqnarray}
A(r) & \approx & 1+\left[ {1\over 2}\kappa^2\phi_0^2
+{1\over 15}(\kappa^2V(0) +\Lambda) + {\bar{R}^{(4)} \over 6}
\right] r^2 \,,\nonumber\\
B(r) &\approx& 1 + \left[ -{2\over 15}(\kappa^2V(0)+\Lambda)
+{\bar{R}^{(4)}\over 12}\right] r^2\,.
\label{eq=BCAB}
\end{eqnarray}
Here, $\phi_0$ is determined by Eq.~(\ref{eq=phi}),
\begin{equation}
\kappa^2\phi_0^2 = {2\over 3}\lambda\eta^2
-{22\over 45}[\kappa^2V(0)+\Lambda ]-{\bar{R}^{(4)} \over 9}\,,
\end{equation}
and $B$ is arbitrary up to a constant multiplication. We choose
this constant so that $B(0)=1$. 

Asymptotics at large $r$ depend on our choice of $\Lambda$.  For
$\Lambda =0$, the asymptotic geometry at large $r$ is conical.  If the
symmetry-breaking scale exceeds the Planck scale, $\kappa\eta \gtrsim
1$, the geometry possesses a coordinate singularity at a finite
distance from the monopole core.  For $\Lambda >0$, the
asymptotic geometry mimics de Sitter spacetime in the higher
dimensions, with a coordinate singularity corresponding to
the de Sitter horizon.  For $\Lambda <0$, the case we present,
the geometry approaches that of anti-de Sitter space asymptotically.
As in the $\Lambda =0$ case, the geometry of super-Planckian
monopoles possesses a coordinate singularity close to its core.
More varieties of the geometry for extra dimensional global defects
were discussed in Ref.~\cite{Vilenkin}.  

We examine the $\Lambda < 0$ case quantitatively. Asymptotically, 
for $\phi\approx \eta$ and $r\gtrsim \delta_\Lambda$, where
\begin{equation}
\delta_\Lambda \sim {\kappa\eta \over \sqrt{|\Lambda |}}\,,
\end{equation}
the local curvature ${\cal R}$ becomes dominated by the cosmological 
constant $\Lambda$, rather than the monopole stress energy. 
The Einstein equations (\ref{einsteqns}) then also have sources 
dominated by  $\Lambda$. They give asymptotic solutions for the metric 
\begin{eqnarray}
A &\approx & ar^{-2} = - \frac{15}{\Lambda r^2}\nonumber\\
B& \approx & br^2
\end{eqnarray}
under the ansatz $A = cB^{-1}$. Note that the coefficient for $A$ is
fixed, giving the correct signature only when $\Lambda$ is negative. 
The coefficient $b$ is unrestricted by the Einstein
equations; however, our choice to set $B = 1$ at the origin determines $b$.    
For this asymptotic metric, the scalar-field equation (\ref{eq=phi}) yields 
asymptotic form for the monopole $\phi$ as 
\begin{equation}
\phi(r) = \eta \left[ 1 - \frac{15}{4|\Lambda|r^2} + O(r^{-4}) \right]\,.
\end{equation}

We display numerical solutions for $A$ and $B$ over the whole
range of $r$ in Fig.~\ref{fig=AB}. The appropriate asymptotic
behavior is noted.

From now on we focus on inducing a four dimensional universe 
on the brane which is like our own. 
The seven dimensional theory induces an effective four dimensional theory, 
with Einstein equations
\begin{equation}
\bar{G}^{\mu}_{\nu} = \bar{R}^{\mu}_{\nu} -{1\over 2} 
\delta^{\mu}_{\nu}\ {\bar{R}^{(4)}} 
=-\Lambda^{(4)} \delta^{\mu}_{\nu}\,.
\label{effeinst}
\end{equation}
Here, barred terms are those induced by the
apparent 4D metric $\bar{g}_{\mu\nu}$, and
$\Lambda^{(4)}$ is the apparent 4D cosmological
constant, which includes contributions from the 7D
cosmological constant, the 7D curvature ${\cal R}$, and
the monopole stress-energy tensor. No $T_{\mu\nu}$ term appears, as we
absorb the monopole's stress energy, diagonal in the longitudinal
brane dimensions, into the effective cosmological constant
$\Lambda^{(4)}$. Note that our full Einstein equations (\ref{einsteqns}), 
which  told us $\bar{G}^\mu_\nu = - \bar{R}^{(4)}/4\ \delta^\mu_\nu$, 
can be read   as an effective equation for $\bar{G}^\mu_\nu$, 
with effective 4D cosmological constant 
\begin{equation}
\Lambda^{(4)} = {B\over A}\left[ -{3\over 2}{B'' \over B} 
+{3\over 4}{A'B' \over AB}
-3{B' \over Br} +{A' \over Ar}
+{A-1 \over r^2}\right] - B 
\kappa^2\left[ {\phi'^2 \over 2A} +{\phi^2 \over r^2}+ V(\phi)
\right] -B\Lambda \,.
\end{equation}
However, this detailed relationship need not be exploited. 
Henceforward we take the 4D
background metric flat, $\bar{R}^{(4)}=0$. This results in apparent 4D
Minkowski space, $\bar{g}_{\mu\nu} = \eta_{\mu\nu}$, with the numerical 
and asymptotic solutions to Einstein's equations shown. 
These solutions automatically have vanishing 4D 
cosmological constant $\Lambda^{(4)}$, 
because  Einstein's equations (\ref{einsteqns}) are equivalent 
to the effective 4D equations (\ref{effeinst}), 
which require $\Lambda^{(4)} = 0$ when $\bar{g}_{\mu\nu} = \eta_{\mu\nu}$. 

Before we close this section, let us comment on the
physical parameters in the model. 
The 7D Planck mass, $M_N$, 
we take to be at the particle scale of TeV; 
our model solves the hierarchy problem by inducing from this natural 
$M_N$ the large effective 4D Planck mass $M_{Pl}$. 
This leaves three free parameters, $\lambda$, $\eta$, and
$\Lambda$.
The self-coupling constant $\lambda$ is always accompanied by
$\eta$ to give the mass of the monopole field, $m_m = \sqrt{\lambda}\eta$.
Since this mass is seven dimensional, we expect its upper bound
does not exceed TeV scale.
The symmetry-breaking scale $\eta$ is not limited; 
however, we confine ourselves to sub-Planckian monopoles ($\kappa\eta < 1$), 
which remain nonsingular at all $r$.
$\Lambda$, which is negative for our solution, can
have arbitrary magnitude.

\section{Graviton zero mode and Planck mass}
\label{sec=MP}

In this section we discuss localization of the graviton
zero mode.
We assume a small perturbation $h_{MN}$ to the background metric whose 
4D part $\bar{g}_{\mu\nu}$ is flat. Then the metric becomes
\begin{eqnarray}
ds^2  &=& [B(r)\eta_{\mu\nu}+h_{\mu\nu}]dx^\mu dx^\nu
	+A(r)dr^2 +r^2d\Omega^2\,,\nonumber\\
&=& B(r)(\eta_{\mu\nu}+\bar{h}_{\mu\nu})dx^\mu dx^\nu
	+A(r)dr^2 +r^2d\Omega^2\,.\label{eq=hbar}
\end{eqnarray}
Here, $\bar{h}_{\mu\nu}$ gives the apparent 4D graviton field.
We assume the only non-zero components of the perturbation
are $h_{\mu\nu}$($h_{\mu j}=h_{ij}=0$). 
We also apply transverse-traceless gauge on $h_{MN}$,
$h_M^M=0$ and ${h_{MN|}}^N=0$, where the vertical bar $|$ 
in the subscript denotes the covariant derivative with respect to
the background metric.
Einstein's  equations for $h_{MN}$ give  
\begin{equation}
{h_{MN|A}}^A +2R^{(B)}_{MANB}h^{AB} -2R^{(B)}_{A(M}h_{N)}^A
-{4\over 2-n}\Lambda h_{MN} -8\pi G_N{4\over 2-n}Vh_{MN}=0\,.
\end{equation}
Here, the superscript (B) represents that the quantity is 
evaluated in the unperturbed background metric $g^{(B)}_{MN}$. 
With given gauge conditions on $h_{MN}$ all terms but the 
first two cancel. Therefore, the equation
for $h_{MN}$ reduces to that in the curved vacuum background,
\begin{equation}
{h_{MN|A}}^A +2R^{(B)}_{MANB}h^{AB} =0\,.
\label{eq=hmn}
\end{equation}

Equation~(\ref{eq=hmn}) is equivalent for every component of
$h_{\mu\nu}$. We thus seek solutions of the form  
$h_{\mu\nu}=\hat{e}_{\mu\nu}h$. 
Equation~(\ref{eq=hmn}) reduces to
\begin{eqnarray}
{1\over A}{\partial^2\over \partial r^2} h
&-& {1\over A}\left( {1\over 2}{A'\over A}-{2\over r}\right)
{\partial \over \partial r}h
+ {1\over r^2} \left( {\partial^2 \over \partial \theta^2}
+{\cos\theta \over \sin\theta}{\partial \over\partial\theta}
+{1 \over \sin^2\theta}{\partial^2 \over\partial\varphi^2}
\right) h \nonumber\\
{}&-& {1\over A}\left( {B''\over B} - {1\over 2}{A'B'\over AB}
+2{B'\over Br}\right) h +{1\over B}\Box^{(4)} h = 0\,,
\label{eq=h}
\end{eqnarray}
where $\Box^{(4)}$ is the d'Alembertian in flat 4D Minkowski space.
We separate variables, taking 
$h=e^{ip_\mu x^\mu}R(r)\Theta (\theta)\Phi (\varphi)$. 
Here, $\Box^{(4)}h = -p_\mu p^\mu h = m^2h$ determines the apparent 
4D graviton mass $m$. The angular equation
\begin{equation}
{1\over \Theta}\left(
{\partial^2\Theta \over \partial\theta^2} 
+{\cos\theta \over \sin\theta}
{\partial\Theta \over \partial\theta}\right)
+{1\over \sin^2\theta}{1\over \Phi}
{\partial^2\Phi \over\partial\varphi^2}
\equiv -l(l+1)\,,
\end{equation}
is solved by  spherical harmonics, 
$\Theta (\theta)\Phi (\varphi) = Y_{lm}(\theta ,\varphi)$. 
This leaves the radial wave function, $R(r)$. Because the apparent 
four dimensional  graviton is described
by $\bar{h}_{\mu\nu} = h_{\mu\nu}/B(r)$ [from  Eq.~(\ref{eq=hbar})], 
we solve instead for its wave function, $\bar{R}(r) = R(r)/B(r)$.
From Eq.~(\ref{eq=h}),  this  satisfies 
\begin{equation}
-B \Box^{(B)} \bar{R}(r)  
= \left[
-{B\over A}{d^2 \over  dr^2}
+{B\over A}\left( {1\over 2}{A'\over A}
-2{B'\over B}-{2\over r}\right){d \over dr}
+{l(l+1)B \over r^2}
\right] \bar{R}(r)  
=  m^2\bar{R}(r)\,.
\label{eq=Req}
\end{equation}

To determine allowed solutions and boundary conditions, 
we rewrite equation (\ref{eq=Req}) in Sturm-Liouville form:
\begin{equation}
\frac{1}{A^{1/2}Br^2} \frac{d}{dr}
\left( A^{-{1/2}}B^2r^2 \frac{d\bar{R}}{dr} \right)  
+\left[ m^2 - {l(l+1)B \over r^2} \right] \bar{R}  = 0\, .
\label{eq=ReqSL}
\end{equation}
This is a Sturm-Liouville equation singular at both the origin 
and infinity, where regular boundary conditions must hold:
\begin{equation} 
\text{$\bar{R}$ bounded},\quad 
A^{-{1/2}}B^2r^2 \bar{R} \bar{R}' \to 0\,, \quad
\text{for $r\to 0, \infty$\,.}
\label{SLbc}
\end{equation}
We seek solutions $\bar{R}$ for allowed values of the eigenvalue  
$m^2$; for simplicity, we restrict ourselves to the  $l=0$ case for gravitons.

We consider first the graviton zero mode, $m=0$.
In this case Eq.~(\ref{eq=Req}) is integrable, with  solution
\begin{equation}
\bar{R}_0(r) = \bar{R}_0 
+\bar{R}_1 \int^r_0 {\sqrt{A} \over B^2r^2} dr\,,
\label{eq=barR0}
\end{equation}
where $\bar{R}_0$ and $\bar{R}_1$ are constants.
At $r\approx 0$, $A\approx 1$ and $B\approx 1$. The second term
in the solution becomes proportional to $1/r$.
It is irregular at $r=0$. Even though it is normalizable,
we exclude this  term because of its irregular
behavior at the origin.
Then the only possible solution is $\bar{R}(r) = \bar{R}_0$
which is a constant. This solution is regular
everywhere, but is not normalizable. 
From the Sturm-Liouville form of the radial equation (\ref{eq=ReqSL}), 
the normalization weight for this zero mode solution is $A^{1/2}B r^2$. 
Thus the graviton zero mode has normalization
\begin{equation}
\bar{R}_0^2 \int dr r^2B\sqrt{A}\, ,
\label{eq=normzero}
\end{equation}
which  diverges. 
Physically, the origin of this divergence lies in the infinite  volume of
the extra dimensions.

This normalization of the zero mode is directly related to
the localization of 4D gravity on the brane.
When the zero mode is non-normalizable, 
4D gravity cannot be realized, as the induced  
4D Planck mass on the brane becomes infinite.

This effective 4D Planck mass can be calculated directly. 
For general $\bar{g}_{\mu\nu}$, our seven dimensional theory 
has in its action the  gravitational term
\begin{eqnarray}
{\cal S}_g &=& \frac{1}{16\pi G_N} \int d^7x \sqrt{-g}{{\cal R}}
\nonumber\\
{} &=& {1 \over 16\pi G_N} \int d^4x\sqrt{-\bar{g}} 
\int drd\theta d\varphi r^2\sin\theta B^2\sqrt{A}
\left[ {\bar{R}^{(4)} \over B} + {\cal R} (\bar{g}_{\mu\nu} = \eta_{\mu\nu})
\right] 
\,.
\end{eqnarray}
Here, we have evaluated the 7D Ricci scalar ${\cal R}$, in terms of
$\bar{R}^{(4)}$, the 4D Ricci scalar induced by $\bar{g}_{\mu\nu}$,
and ${\cal R}\ (\bar{g}_{\mu\nu} = \eta_{\mu\nu})$, the 7D Ricci
scalar, evaluated when $\bar{g}_{\mu\nu}$ is the flat 4D Minkowski
metric. Viewing the $\bar{R}^{(4)}$ term as an effective 4D
gravitational action, we read off  its effective coupling constant
\begin{equation}
M_{Pl}^2 = 4\pi M_N^{2+n}\int dr r^2B\sqrt{A}\,,
\label{eq=pm}
\end{equation}
after the substitution $G_N=1/8\pi M_N^{2+n}$.
The integral in Eq.~(\ref{eq=pm}) is directly proportional to
the normalization of the graviton zero mode in Eq.~(\ref{eq=normzero}).
Again, this integration is divergent, making the 4D Planck mass
infinite.

In the models of $n\leq 2$, the integral of Eq.~(\ref{eq=pm}) is 
finite due to the exponentially decreasing warp factor.
However, in our model, 
the warp factor is $B\sim r^2$ at large $r$ and the integral diverges.
Therefore, in order to have normalizable graviton zero mode
and  finite 4D Planck mass $M_{Pl}\sim 10^{18}$GeV, 
we introduce a cut-off.
Such a cut-off should arise dynamically, due to formation of  other topological defects
around the monopole. It can be imposed formally, by
considering a monopole and anti-monopole pair.

We estimate the cut-off radius $r_*$ inducing the observed hierarchy; 
that is, inducing 4D $M_{Pl}\sim 10^{18}$GeV from a 
7D Planck mass  of TeV scale, $M_N\sim$TeV.
The main contribution to the integral in Eq.~(\ref{eq=pm})
comes from the region outside the core where $r^2B\sqrt{A}$
is much larger than unity.
In this region, from the result of the previous section, 
$B\sqrt{A}=b\sqrt{a}r$.
Then Eq.~(\ref{eq=pm}) becomes
\begin{equation}
M_{Pl}^2 = 4\pi b\sqrt{a}M_N^{2+n}\int^{r_*}dr r^3
	\simeq \pi b\sqrt{a}M_N^5r_*^4\,.
\end{equation}
This gives the cut-off radius 
\begin{equation}
r_* = \left( {\sqrt{|\Lambda |}M_N \over \sqrt{15}\pi b}\right)^{1\over 4}
\left( {M_{Pl} \over M_N} \right)^{1\over 2} {1\over M_N}
\simeq \left( {\text{TeV} \over b|\Lambda |^{-1/2}} \right)^{1\over 4} 10^{-10}\rm{cm}
\,,
\end{equation}
where $b|\Lambda |^{-1/2}$ is a mass scale, 
undetermined because $\Lambda$ is a free parameter. 
However, a range of allowed parameters yield a cut-off radius $r_* $
acceptable by  current observation. 
Regarding $r_* $  as the size of the extra dimensions
in our model, gravity takes its usual 4D form so long as the separation of
probe particles in the brane remains larger than $r_* $.
Therefore, the cut-off radius $r_* $ cannot not exceed 1 mm, 
the current observational resolution of gravitational measurement.

\section{Massive Kaluza-Klein modes}

In this section we discuss the massive Kaluza-Klein gravitons and
their correction to the Newtonian gravitational potential on
the brane.

Massive gravitons obey the radial graviton equation~(\ref{eq=ReqSL}),
with boundary conditions~(\ref{SLbc}). This Sturm-Liouville problem
for the 4D graviton wave function $\bar{R}$ leads directly to two
conclusions: first, the existence of the zero-mode solution examined
in the previous section; and second, by standard Sturm-Liouville
arguments, to positive definiteness of the allowed mass eigenvalues,
$m^2 \ge 0$, with the regular boundary conditions~(\ref{SLbc}). 
In this form we thus see that our theory allows no tachyonic graviton modes.

To find the allowed  mass eigenvalues, however, we 
recast the radial equation~(\ref{eq=ReqSL}) as a 
non-relativistic Schr\"odinger-type equation.
We rescale both the radial coordinate and the
wave function by 
\begin{eqnarray}
z &=& \int^r_0 dr \sqrt{{A\over 2B}}\,,
\label{eq=z}\\
\psi (z) &=& NB^{3\over 4}
{\bar{R}(r)r \over z}\,,
\label{eq=psi}
\end{eqnarray}
where $N$ is a constant. 
Then the radial equation becomes
\begin{equation}
\left[ -{1\over 2}{d^2\over dz^2} -{1\over z}{d \over dz}
+U(z)\right] \psi (z) = m^2\psi (z)\,,
\label{eq=psiflat}
\end{equation}
with potential 
\begin{equation}
U(z) = {B\over A}\left[ {3 \over 4}{B''\over B}
+{3\over 16}\left( {B'\over B}\right)^2
-{3\over 8}{A'B' \over AB}+2{B'\over Br}
-{1\over 2}{A' \over Ar}\right]\,.
\label{Veff}
\end{equation}
Here, the prime denotes  differentiation with respect to $r$.
Equation~(\ref{eq=psiflat}) is in the form of a Schr\"odinger-type equation 
in three dimensions, with spherically symmetric effective potential $U$ 
and energy eigenvalue $m^2$.
Therefore, the shape of the effective potential provides
insight both into the allowed mass spectrum, and into
how gravitons are confined in space.

However, the boundary conditions are nontrivial. Note that we must
reproduce the boundary conditions (\ref{SLbc}) of the original
Sturm-Liouville problem for $\bar{R}$, in order to reproduce the proper  mass
spectrum (eigenvalue spectrum) $m^2$. These boundary conditions occur at 
$r=0$, where $z=0$, and at $r\rightarrow \infty$, 
where $z$ takes on the finite positive value
\begin{equation}
{\cal Z}\equiv \int^\infty_0 dr \sqrt{{A\over 2B}}\,.
\end{equation}
This is finite because the integrand falls as $1/r^2$ at large $r$.
These boundary conditions require that $\bar{R}$, related to $\psi$ by 
Eq.~(\ref{eq=psi}), must remain bounded 
at $z=0$, or $z={\cal Z}$, 
and that
\begin{eqnarray} 
0 &=& A^{-{1/2}}B^2r^2 \bar{R}\bar{R}'\nonumber \\
&=& z^2A^{-{1/2}}B^{1/2}\psi^2  \left( \frac{z'}{z} -\frac{3}{4} \frac{B'}{B} 
- \frac{1}{r} + \frac{\psi'}{\psi}  \right) \,,
\label{psiregbc}
\end{eqnarray}
where the prime again denotes differentiation with respect to $r$.
We note that the quantity in parentheses vanishes automatically 
for the zero mode, for which $ \bar{R}' = 0$, and for no other solution. 
For all other modes, this quantity scales as $1/r$, 
leading to effective regularity conditions 
\begin{equation}
\frac{z^2}{r} \sqrt{\frac{B}{A}} \psi^2  \to 0\,, \quad
z \to 0\,, {\cal Z}\,. 
\label{psireg}
\end{equation}

With these boundary conditions in mind, we consider our Schr\"odinger equation 
(\ref{eq=psiflat}) with  effective potential (\ref{Veff}). 
Since we know the functional form for $A$ and $B$ at $r \simeq 0$
and $r \gg \delta_\Lambda$, 
let us evaluate the functional form of $U(z)$ and
the wave function in these two asymptotic regions.
The numerical plot of $U(z)$ for the whole range is shown in
Fig.\ref{fig=U}.

For small  $r$, with the aid of $A$ and $B$ in
Eq.~(\ref{eq=BCAB}),  we approximate
$z =\int^r_0 dr\sqrt{A/2B}\approx r/\sqrt{2}$. 
The potential reads
\begin{equation}
U(z) = -\left[ {1\over 2}\kappa^2\phi_0^2
+{4\over 5}(\kappa^2V(0)+\Lambda)\right] + {\cal O}(z^2)\,.
\label{eq=Usmall}
\end{equation}
This constant potential must automatically lay below zero, 
as we know the zero mode, with ``energy eigenvalue'' $m^2 = 0$, 
remains a solution to the transformed Schr\"odinger 
equation~(\ref{eq=psiflat}). 
Therefore, the wave function $\psi (z)$ approaches 
the plane-wave solution for the given eigenvalue $m$. 
To the lowest-order correction to $U(z)$,
the solution to the wave equation~(\ref{eq=psiflat}) becomes
\begin{equation}
\psi_m(z) = c_1 {\sin\sqrt{2(m^2+U_0)}z \over z}
+c_2 {\cos\sqrt{2(m^2+U_0)}z \over z}\,,
\end{equation}
where $U_0 \equiv -U(0) = {1\over 2}\kappa^2\phi_0^2
+{4 \over 5}[\kappa^2 V(0) +\Lambda]$. 
The cosine part is irregular at $z=0$, 
where $\bar{R} \sim \psi$ becomes unbounded. 
We thus exclude it, setting $c_2=0$.

At large distances
\begin{equation}
z=\int^r_0 dr \sqrt{{A\over 2B}} \approx {\cal Z} - \sqrt{{a\over 2b}} 
{1\over r}\,.
\end{equation}
The potential in this region becomes
\begin{equation}
U \approx {35b\over 4a}r^2 = {K \over ({\cal Z} -z)^2}\,,
\label{eq=Ularge}
\end{equation}
where $K\approx 35/8$. 
As $z\to {\cal Z}$ ($r\to\infty$), $U$ goes to infinity.
The solution to the wave equation with this $U(z)$ is given by
\begin{equation}
\psi_m(z) = c_1 {\sqrt{{\cal Z}-z} \over z}
J_{{1\over 2}\sqrt{1+8K}}[\sqrt{2}m({\cal Z}-z)]
+c_2 {\sqrt{{\cal Z}-z} \over z} 
Y_{{1\over 2}\sqrt{1+8K}}[\sqrt{2}m({\cal Z}-z)]\,,
\end{equation}
where ${1\over 2}\sqrt{1+8K} \approx 3$.
$J$ ($Y$) is the Bessel function of the first (second) kind.
The $Y$-function becomes irregular as $z\to {\cal Z}$, 
violating our nontrivial effective regularity condition~(\ref{psireg}). 
We thus exclude it, $c_2=0$.
From the two asymptotic behaviors of the wave function and the
shape of the potential, it is clear that the wave function for
the massive mode is normalizable regardless of the cut-off.

Now we are in a position to evaluate the mass spectrum.
Equation~(\ref{eq=psiflat}) is in the form of a non-relativistic
Schr\"odinger equation, whose binding potential suggests 
a discrete spectrum of eigenvalues.
More strictly, as a 
Sturm-Liouville equation over the finite variable $z=[0,{\cal Z})$, 
this equation must yield discrete eigenvalues $m^2$. 
We know, from our problem statement in radial variables, 
that the smallest eigenvalue is $m^2 = 0$, 
with eigenfunction automatically obeying the regular 
boundary conditions (\ref{psiregbc}).

Quantization of the spectrum is usually acquired by imposing
regular boundary conditions on the excited eigenfunction
solutions. Here, however, we know only these solutions' asymptotic
behaviors, and  thus cannot impose both boundary conditions on a
single candidate solution. That is, while we may discard irregular
asymptotic behavior, we have no guarantee that an eigenfunction
regular at the origin continues to one that is regular also at $z =
{\cal Z}$. Yet it is precisely the special case where this
continuation remains regular, which determines the allowed eigenvalues
$m^2$.

To investigate the eigenvalue spectrum, then, 
we must approximate our equation~(\ref{eq=psiflat}) 
in such a way that a single closed-form solution connects 
the two asymptotic regions $z\sim 0$ and $z\sim {\cal Z}$.  
We achieve this by approximating our potential  $U(z)$ by
\begin{equation}
U(z) = -U_0 -{K \over {\cal Z}^2} + {K \over ({\cal Z}-z)^2}\,.
\label{eq=Uapprox}
\end{equation}
This potential approximates Eqs.~(\ref{eq=Usmall}) and
(\ref{eq=Ularge}) in the two asymptotic regions ($z\to 0$ and $z\to
{\cal Z}$). The intermediate region is also well approximated.  Even
though this approximate potential is not exactly the same as the
original one, we can estimate the mass eigenvalues in a good
approximation. Further corrections could be found
numerically by the variational method, for a specific parameter
choice.  

With this approximate $U$ the solution of the graviton
equation (\ref{eq=psiflat}) is
\begin{eqnarray}
\psi_m(z) &=& C_1{\sqrt{{\cal Z}-z} \over z}
J_{{1\over 2}\sqrt{1+8K}} \left[ \sqrt{2(m^2+U_0+{K\over {\cal Z}^2})}
({\cal Z}-z) \right]\nonumber\\
{} &+& C_2{\sqrt{{\cal Z}-z} \over z}
Y_{{1\over 2}\sqrt{1+8K}} \left[ \sqrt{2(m^2+U_0+{K\over {\cal Z}^2})}
({\cal Z}-z) \right]\,.
\label{eq=psiapprox}
\end{eqnarray}
Imposing our effective regularity condition~(\ref{psireg})
at $z\to {\cal Z}$ excludes the $Y$-solution,
$C_2 =0$. 
At $z\to 0$, it forces the 
$J$-solution to vanish. Thus the argument of $J$
at $z=0$ must be a zero of the Bessel function,
\begin{equation}
\sqrt{2(m^2+U_0+{K\over {\cal Z}^2})}{\cal Z} = x_{3,j} \simeq \pi (j+{5\over 4})\,,
\quad j=1,2,...\,. 
\label{eq=xj}
\end{equation}
Here, $x_{3,j}$'s are the zeros of $J_3(x)$ and $\pi (j+{5\over 4})$ is a very
good approximation even for $j=1$.
The discrete mass spectrum is then given by
\begin{equation}
m_j^2 = -U_0 -{K\over {\cal Z}^2} 
+ {\pi^2\over 2{\cal Z}^2}\left(j+{5\over 4}\right)^2\,.
\label{eq=mj}
\end{equation}
The mass gap between the adjacent modes is
\begin{equation}
\Delta m_j^2 = {\pi^2 \over {\cal Z}^2}\left( j+ {7\over 4} \right)\,.
\label{eq=mgap}
\end{equation}
The value of $m_j$ depends on  ${\cal Z}$ which is given by
\begin{equation}
{\cal Z} = z(r\to\infty ) =\int^\infty_0dr\sqrt{A\over 2B}
 \approx \int^{\delta_\Lambda}_0 dr\sqrt{A\over 2B}
+{1\over \delta_\Lambda}\sqrt{a\over 2b}\,.
\end{equation}
The last approximation comes after we separate the integration
at $r=\delta_\Lambda$ beyond which we know the asymptotic
$A$ and $B$. 
From the above description it is clear that ${\cal Z}$ depends much on
the parameters of the model.
The estimation of ${\cal Z}$ is not quite possible because of the
integration over the inner region.

We note that this spectrum $m_j^2$ describes the excited Kaluza-Klein 
modes, with $m_j^2 > 0$. 
The exact $m^2=0$ mode~(\ref{eq=barR0}) does not appear
as a solution to our approximate potential~(\ref{eq=Uapprox}),
nor does it obey the
imposed effective regularity conditions~(\ref{psireg}); 
although it automatically
obeys the more general regularity conditions (\ref{psiregbc})
transcribed directly from the radial graviton problem,
Eq.~(\ref{eq=ReqSL}). $m_j^2$, with $j=1$, is then the first Kaluza-Klein
mode obeying the excited boundary conditions~(\ref{psireg}) , and is
thus guaranteed by the Sturm-Liouville form of the original radial
problem Eq.~(\ref{eq=ReqSL}) to be positive. Its positiveness is not
apparent from its form (\ref{eq=mj}); however, self-consistency of the
eigenvalue problem requires that the highly derivative quantities
$U_0$ and ${\cal Z}$ must emerge with values such that $m_1^2 >0$.

The correction to the Newtonian potential by the massive 
KK gravitons is given by
\begin{equation}
\Delta V_{Newt}({\bf\text{x}}) = 
\sum_j 
G_NM_1M_2 {e^{-m_j{\bf \text{x}}} \over {\bf \text{x}}} 
|\psi_{m_j}(z=0)|^2 \,,
\end{equation}
where ${\bf \text{x}}$ is the distance between 
the particles of mass $M_1$ 
and $M_2$ in the 4D brane. 
For ${\bf \text{x}}>1/m_j$, 
the correction is exponentially suppressed.
If $m_j$ (especially the lowest mode) is massive enough,
the correction to the Newtonian potential will be
negligable at the observational resolution,
unless the KK-graviton couplings $G_N |\psi_{m_j}(z=0)|^2$
diverge.

With the wave function~(\ref{eq=psiapprox}) 
we can evaluate 
these couplings of KK gravitons to the brane
located at the center of the monopole, $z=0$.
The normalization of the $l=0$  wave function~(\ref{eq=psiapprox}),
$4\pi\int^{\cal Z}_0 |\psi_{m_j}(z)|^2z^2dz = 1$,
gives the normalization constant
\begin{equation}
C_1^2(m_j) = {1 \over 2\pi {\cal Z}^2
[J_3^2(x_{3,j})-J_2(x_{3,j})J_4(x_{3,j})]}
=-{1 \over 2\pi {\cal Z}^2 J_2(x_{3,j})J_4(x_{3,j})}\,.
\end{equation}
Here, we used the relation~(\ref{eq=xj}) for $x_{3,j}$ and
its definition, $J_3(x_{3,j})=0$.
Then the coupling reads
\begin{eqnarray}
G_N |\psi_{m_j}(z=0)|^2 &=&
G_N C_1^2(m_j) \lim_{z\to 0}
\left| {\sqrt{{\cal Z}-z}\over z}
J_3\left[{x_{3,j}\over {\cal Z}}({\cal Z}-z)\right]
\right|^2 \nonumber\\
{}&=& -{G_N\over 2\pi {\cal Z}^3}
{
\left[ (5/2) J_3(x_{3,j})
-x_{3,j} J_2(x_{3,j}) \right]^2
\over
J_2(x_{3,j})J_4(x_{3,j})
} \nonumber\\
{} &=& {G_N\ \over 2\pi{\cal Z}^3}
x_{3,j}^2\,,
\label{eq=KKcoupling}
\end{eqnarray}
where we used the recurrence formula
$J_4(x)= 6J_3(x)/x -J_2(x)$.

Since $m_1$ and ${\cal Z}$ emerge as such 
parameter-dependent quantities, let us consider the
possibility of small mass gap $\Delta m_j$.
For small mass gap, we take the continuum limit, 
$\Delta m_j \to 0$.
The Kaluza-Klein correction to the gravitational potential 
is then
\begin{equation}
\Delta V_{Newt}({\bf \text{x}}) =
4\pi \int_0^\infty d\left( {m\over e_m}\right)
\left( {m\over e_m}\right)^2G_NM_1M_2
{e^{-m{\bf \text{x}}} \over {\bf \text{x}}}
|\psi_m(0)|^2\,,
\end{equation} 
where the additional $m$ factor comes from the extra
dimensional plane-wave continuum density of states,
and $e_m$ is the unit mass.
From Eqs.~(\ref{eq=xj}) and (\ref{eq=KKcoupling})
this correction becomes
\begin{equation}
\Delta V_{Newt}({\bf \text{x}}) =
G_NM_1M_2 {1\over e_m^3} \left[ {8\over {\cal Z}}
\left( U_0+{K \over {\cal Z}^2} \right)
{1\over {\bf \text{x}}^4}
+{96 \over {\cal Z}}{1\over {\bf \text{x}}^6}\right]\,.
\end{equation}
This correction negligibly affects the $1/{\bf \text{x}}$
behavior of the Newtonian potential in four dimensions.

\section{Matter fields}

In this section we investigate the effective 
field theory induced on the brane for 7D matter fields. 
Either Dirac or Klein-Gordon scalar fields obey a
7D Klein-Gordon field equation:
\begin{equation}
\Box^{(7)} \phi_{KG} =
{1\over \sqrt{-g}}\partial_A (\sqrt{-g}g^{AB}\partial_B )
\phi_{KG} = M_{KG}^2\phi_{KG}\,.
\end{equation}
where $\phi_{KG}$ and $M_{KG}$ are the 7D  field and
its mass.
To induce an effective 4D field, we separate variables, 
\begin{equation}
\phi_{KG} = e^{ip_\mu x^\mu} R_{KG}(r) Y_{lm} (\theta, \phi )\,,
\end{equation}
with  radial equation
\begin{equation}
\left[
-{B\over A}{d^2 \over  dr^2}
+{B\over A}\left( {1\over 2}{A'\over A}
-2{B'\over B}-{2\over r}\right){d \over dr}
+{l(l+1)B \over r^2} +M_{KG}^2B
\right] R_{KG}(r) = m_{KG}^2R_{KG}(r)\,,
\nonumber
\end{equation}
where $\Box^{(4)}\phi_{KG} = m_{KG}^2\phi_{KG} $ 
determines the effective 4D mass. 
This eigenvalue equation determines the induced 4D mass of 
the scalar field and its Kaluza-Klein excitations.

The radial equation is exactly the same as 
equation~(\ref{eq=Req}) for gravitons, with one more term
coming from the 7D mass. 
Due to this additional term, for $M_{KG}$ nonzero, 
no zero-mode type solution exists. 
However, we may recast this equation
to the Schr\"odinger-type 
equation~(\ref{eq=psiflat}) by the same 
transformations (\ref{eq=z}) and (\ref{eq=psi}).
The potential for the KG-wave equation is then given by
\begin{equation}
U_{KG}(z) = U(z) +M_{KG}^2B\,.
\end{equation}
Therefore, the potential is corrected from the graviton
potential by adding a term with the 7D mass.
At small $r$  it asymptotes to
\begin{equation}
U_{KG}(z) = M_{KG}^2 -U_0 +{\cal O}(z^2)
\equiv -U_0^{KG} +{\cal O}(z^2)\,,
\end{equation}
and at large $r$ it goes to
\begin{equation}
U_{KG}(z) = U(z) +{aM_{KG}^2/2 \over ({\cal Z}-z)^2}
=  {K+ aM_{KG}^2/2 \over ({\cal Z}-z)^2}
\equiv {K_{KG} \over ({\cal Z} -z)^2}\,.
\end{equation}
Then the properties of the potential and the wave function
are the same with those for massive gravitons after we
substitute two constants $U_0$ and $K$ with 
$U_0^{KG}$ and $K_{KG}$.
The localization behavior is the same as that of 
massive KK gravitons. 
Therefore, we have a 4D KG field localized on the brane, 
with discrete mass spectrum
$m_{KG,j}^2 = m_j^2 +(1-a/2{\cal Z}^2)M_{KG}^2$, for $j \ge 1$.

\section{conclusions}

We investigated the brane physics induced by a global monopole 
formed in three extra dimensions.
The metric induced by the monopole background, with negative  cosmological
constant, makes the volume of the extra dimensions infinite.
As usual in models of infinite extra dimensional volume,
the graviton zero mode is not normalizable without a cut-off.
We view a cut-off as physically natural, 
due to formation of adjacent defects during 
the dimension-reducing phase transition, and choose 
the cut-off radius to induce a hierarchical  4D Planck 
mass ($\sim 10^{18}$GeV) from a unified 7D Planck 
mass ($\sim$TeV). 
We have a possible range of parameters to make
this cut-off radius (size of extra dimensions) consistent
with current observations.

Our model admits a discrete spectrum of massive 
Kaluza-Klein gravitons.
These massive modes are not harmful to  4D gravity
on the brane, whether the mass gap is large or small.

Recently it was suggested that 4D gravity can be induced
by the massive KK gravitons even when the zero mode is not
normalizable~\cite{Meta}. 
Four dimensional gravity in that scenario is viable in the intermediate
length scale. The induced graviton is meta-stable on the brane
and decays to the bulk at a finite life time. 
Therefore, at a large scale higher
dimensional gravity applies.
The scenario is still under dispute. However, it will be 
challenging to examine its validity in our model 
in which the zero mode
is not naturally normalizable.

In addition to gravitons we found that 7D matter
fields can induce 4D matter  fields
on the brane. The localization picture is very similar
to that of massive gravitons. 
It will be interesting to study the effective field theory 
of matter interactions on the brane, and the resultant
brane cosmology.

\acknowledgments We are grateful to Gia Dvali, Alex Vilenkin, Takahiro
Tanaka, and Gaume Garriga for helpful discussions.  I.C. also
acknowledges Chi-Ok Hwang. This work was supported by the
University Research Committee of Emory University.


\begin{references}

\bibitem{KK}
T. Kaluza, Preus. Acad. Wiss. K {\bf 1}, 966 (1921);
O. Klein, Zeit. Phys. {\bf 37}, 895 (1926).
\bibitem{ADD1}
N. Arkani-Hamed, S. Dimopoulos and G. Dvali,
Phys. Lett. B {\bf 429}, 263 (1998);
Phys. Rev. D {\bf 59}, 086004 (1999).
\bibitem{RS}
L. Randall and R. Sundrum,
Phys. Rev. Lett. {\bf 83}, 3370 (1999);
{\it ibid.} 4690 (1999).
\bibitem{Rubakov2}
V. Rubakov and M. Shaposhnikov,
Phys. Lett. B {\bf 125}, 139 (1983).
\bibitem{Rubakov1}
V. Rubakov and M. Shaposhnikov,
Phys. Lett. B {\bf 125}, 136 (1983).
\bibitem{Shifman}
G. Dvali and M. Shifman,
Phys. Lett. B {\bf 396}, 64 (1997).
\bibitem{Vilenkin}
I. Olasagasti and A. Vilenkin,
Phys. Rev. D {\bf 62}, 044014 (2000).
\bibitem{Cohen}
A. Cohen and D. Kaplan,
Phys. Lett. B {\bf 470}, 52 (1999).
\bibitem{Gregory}
R. Gregory,
Phys. Rev. Lett. {\bf 84}, 2564 (2000).
\bibitem{Shaposhnikov1}
T. Gherghetta and M. Shaposhnikov,
Phys. Rev. Lett. {\bf 85}, 240 (2000).
\bibitem{Shaposhnikov2}
T. Gherghetta, E. Rossel and M. Shaposhnikov,
Phys. Lett. B {\bf 491}, 353 (2000).
\bibitem{Meta}
R. Gregory, V. Rubakov and S. Sibiryakov,
Phys. Rev. Lett {\bf 84}, 5928 (2000);
Phys. Lett B {\bf 489}, 203 (2000);
C. Csaki, J. Erlich and T. Hollowood,
Phys. Rev. Lett. {\bf 84}, 5932 (2000);
Phys. Lett. B {\bf 481}, 107 (2000);
G. Dvali, G. Gabadadze and M. Porrati,
Phys. Lett. B {\bf 484}, 112 (2000);
{\it ibid.} 129 (2000).

\clearpage


\begin{figure}
\begin{center}
\epsfig{file=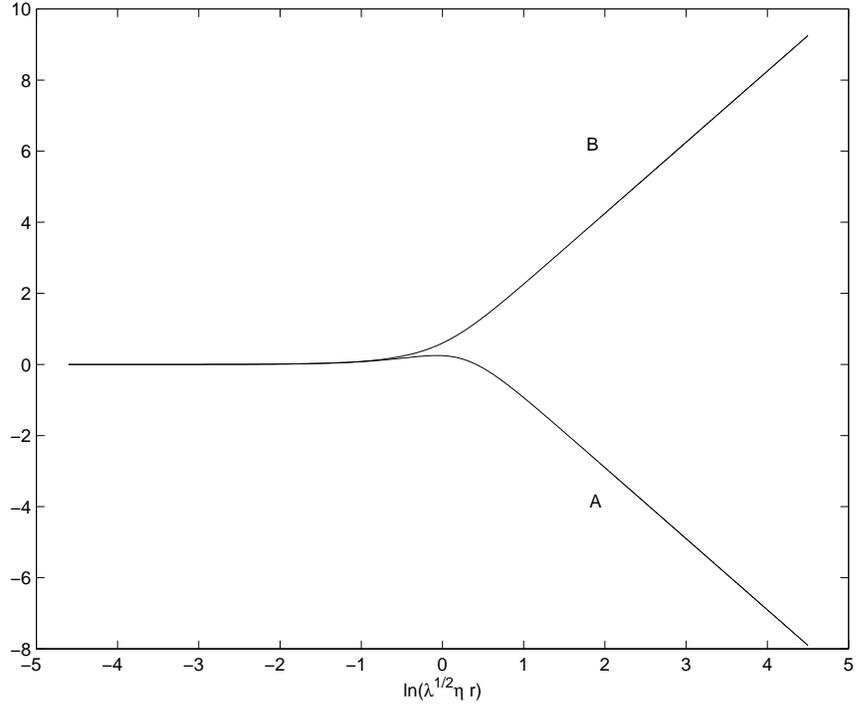,width=4.5in}
\end{center}
\vspace{0.5in}
\caption{
Plot of $A$ and $B$ vs $\sqrt{\lambda}\eta r$ for $\kappa\eta =1$ 
and $\Lambda =-5\lambda\eta^2$.
To see the power-law dependence of $A$ and $B$ on $r$ clearly, 
both axes are shown in logarithmic scale.
This plot shows that $A\sim r^{-2}$ and
$B\sim r^2$ at large $r$.}
\label{fig=AB}
\end{figure}

\clearpage

\begin{figure}
\begin{center}
\epsfig{file=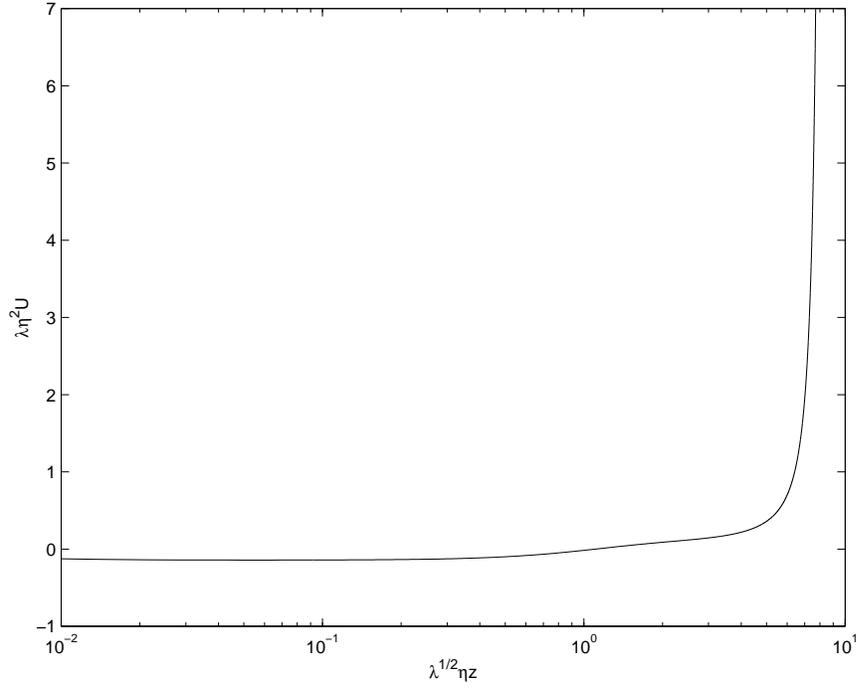,width=4.5in}
\end{center}
\vspace{0.5in}
\caption{
Plot of $U$ vs $z$ for $\kappa\eta =1$ 
and $\Lambda =-0.2\lambda\eta^2$.
To see the central region clearly the x-axis is  in logarithmic scale.
This plot shows that $U$ is constant about the center and
diverges as $z\to {\cal Z}$.}
\label{fig=U}
\end{figure}




\end{references}
\end{document}